\documentstyle[12pt,aaspp4,emulateapj5,psfig]{article}

\def\la{\mathrel{\hbox{\rlap{\hbox{\lower4pt\hbox{$\sim$}}}\hbox{$<$}}}}
\def\ga{\mathrel{\hbox{\rlap{\hbox{\lower4pt\hbox{$\sim$}}}\hbox{$>$}}}}

\font\syvec=cmbsy10                        
\def\bnabla{\hbox{{\syvec\char114}}}       
\def\bphi{\hbox{{\gkvec\char30}}}       
\font\gkvec=cmmib10                         

\lefthead{Agol et al.}
\righthead{Convection in Bright Accretion Disks}
\begin{document}

\title{Two-Dimensional Hydrodynamic Simulations of Convection in 
Radiation-Dominated
Accretion Disks}

\author{Eric Agol (Chandra Fellow)}
\affil{Physics and Astronomy Department, Johns Hopkins University, Baltimore
MD 21218 and Theoretical Astrophysics, Caltech MS 130-33, Pasadena, CA 91125}
\author{Julian Krolik}
\affil{Physics and Astronomy Department, Johns Hopkins University,
    Baltimore, MD 21218}
\author{Neal J. Turner \& James M. Stone}
\affil{Astronomy Department, University of Maryland, 
    College Park, MD 20742}
\begin{abstract}

     The standard equilibrium for radiation-dominated accretion disks has
long been known to be viscously, thermally, and convectively unstable, but
the nonlinear development of these instabilities---hence the actual state of
such disks---has not yet been identified.  By performing local two-dimensional
hydrodynamic simulations of disks, we demonstrate that 
convective motions can release heat sufficiently rapidly as to substantially
alter the vertical structure of the disk.  If the dissipation rate within
a vertical column is proportional to its mass, the disk settles into
a new configuration thinner by a factor of two than the
standard radiation-supported equilibrium.  If, on the other hand, the
vertically-integrated dissipation rate is proportional to the
vertically-integrated total pressure, the disk is subject to the
well-known thermal instability.  Convection, however, biases
the development of this instability toward collapse.  The end result
of such a collapse is a gas pressure-dominated equilibrium at the
original column density.

\end{abstract}


\section{Introduction}

      More than twenty-five years ago, Shakura \& Sunyaev (1973, hereafter SS) 
showed that the inner portions of accretion disks with luminosities near
Eddington are likely to be dominated by radiation
pressure.  Because the vertical component of gravity increases $\propto z$
near the disk midplane, true vertical hydrodynamic equilibrium could
only be achieved if the heating rate is constant with height.
If the physical dissipation rate scales locally with the density, then
hydrostatic and radiative equilibrium require the gas density to be constant 
from the disk midplane to the disk surface.

     Unfortunately, this equilibrium is subject to numerous instabilities.
Lightman \& Eardley (1974) pointed out that if the viscous stress is
proportional to the radiation pressure, perturbations to the surface
density grow on the (comparatively long) viscous inflow timescale.  Shakura \&
Sunyaev (1976) then observed that in these same conditions the thermal
content of the disk is likewise unstable, with a growth time comparable to
the (shorter) thermal timescale.  Bisnovatyi-Kogan \& Blinnikov (1977, hereafter
BKB) noticed that if the radiation is locked to the gas even on short 
length scales
(i.e., if, for the purpose of dynamics, the optical depth is treated as
effectively infinite), such disks should be convectively unstable, for the
specific entropy decreases upward (the radiation pressure decreases upward
while the density is constant); the linear growth rate for convective
``bubbles" was worked out by Lominadze \& Chagelishvili (1984).  This work
has been recently extended by Pietrini \& Krolik (2000), who derived a
hydrodynamic WKB dispersion relation in the presence of rotation and
including the effects of
finite optical depth.  In other recent work, Gammie (1998) demonstrated that
a magnetic field in radiation-supported disks can catalyze a short-wavelength
($kH \gg 1$, where $H$ is the disk scale height) overstable wave mode.
Blaes \& Socrates (2000) found the dispersion relation for
radiation MHD modes within radiation-dominated disks.

    With so many instabilities potentially operating, one must wonder what
is the real state of these disks.  Numerous suggestions exist in
the literature.  
BKB argued that the
disk structure could be determined by imposing the conditions of constant
specific entropy (a state achieved as a result of convection) and
hydrostatic equilibrium.  Liang (1977) defined a new equilibrium by 
parameterizing the efficiency of convective heat transport.  Shakura,
Sunyaev, \& Zilitinkevich
(1978) found an analytic solution for the vertical structure based on 
these same assumptions, but including support from turbulent
motions in the hydrostatic equilibrium.  They were able to find an analytic
solution to the equations granted the assumption that the rms
Mach number of the turbulence took a special value.  Robertson and
Tayler (1981) 
argued that convection does not quench the thermal instability, making rough 
estimates for the effect of convection on transport.  Others (Cannizzo 1992; 
Milsom, Chen \& Taam 1994; Rozanska et al. 1999) have proposed
models based on the ``mixing-length" prescription.  Eggum et al.
(1987), Milsom \& Taam (1997), and Fujita \& Okuda (1998) used 2-D
radiation hydrodynamic simulations to study various aspects of these disks.

     Unfortunately, each of these previous attempts to understand the
structure of radiation-dominated disks has been lacking in one or more
crucial respects.  None of the analytic efforts actually solved the
force equation without making some assumption about the character of the answer,
while none of the numerical simulations had the resolution to actually
determine the internal vertical structure (these simulations were primarily
aimed at exploring issues involving the global behavior of disks rather
than the particulars of vertical structure).

      The object of the work presented here is to use radiation hydrodynamic
simulations to focus on the vertical equilibrium of radiation-dominated disks
so that one may actually determine the structure of bright accretion disks
in the range of radii where they release most of their energy, their innermost
rings.  In addition to the intrinsic interest of this effort for deepening
our understanding of accretion dynamics, solving this question is a
prerequisite for any effort to predict the spectrum of radiation emerging
from these disks---the nature of the disk atmosphere, and therefore the
character of any features imprinted on the spectrum in the disk photosphere,
depends crucially on the vertical distribution of gas density and heat
deposition.  Though the viscous mechanism operating in accretion flows is
likely due to magnetic stresses, we parameterize the effects of viscosity
so as to use a simpler hydrodynamic code.

\section{Simulation Methods}

   The tool we employ is the Zeus simulation code (Stone \& Norman 1992)
with its flux-limited radiation diffusion module 
(Turner \& Stone 2001).  This code solves five coupled partial differential
equations on a fixed Eulerian grid: the mass-continuity equation, the
Navier-Stokes equation, the energy conservation equations for both the
gas and the radiation, and the radiation momentum conservation equation.
Defining the radiation quantities in a cylindrical coordinate frame comoving 
with the fluid, and retaining only those terms first order in $v/c$, these 
equations are:
\begin{equation}\label{eqn:cty}
{D\rho\over D t}+\rho\nabla\cdot{\bf v}=0,
\end{equation}
\begin{eqnarray}\label{eqn:gasmomentum}
\rho{D{\bf v}\over D t} = -\nabla p + {1\over c}\kappa\rho{\bf F} - 
\rho\Omega^2 z {\bf z} - 2\rho\Omega (v_\phi-r_o \Omega) {\bf r} \cr
+ 2\rho\Omega v_r \bphi + 3 \rho\Omega^2 (r-r_o) {\bf r},
\end{eqnarray}
\begin{equation}\label{eqn:radenergy}
\rho{D\over D t}\left({E\over\rho}\right) =
	- \nabla\cdot{\bf F} - \nabla{\bf v}:{\mathsf P}
	+ 4\pi\kappa_P\rho B - c\kappa_E\rho E + \eta,
\end{equation}
\begin{equation}\label{eqn:gasenergy}
\rho{D\over D t}\left({e\over\rho}\right) =
	- p \nabla\cdot{\bf v}
	- 4\pi\kappa_P\rho B + c\kappa_E\rho E,
\end{equation}
and
\begin{equation}\label{eqn:radmomentum}
{\rho\over c^2}{D\over D t}\left({{\bf F}\over\rho}\right) =
	- \nabla\cdot{\mathsf P}
	- {1\over c}\kappa\rho{\bf F}.
\end{equation}
Here the convective derivative $D/Dt\equiv\partial/\partial t + {\bf
v}\cdot\nabla$.  The unit vectors in the radial, azimuthal, and vertical
directions are ${\bf r}$, $\bphi$, and ${\bf z}$, respectively.
The quantities $\rho$, $e$, ${\bf v}$, and $p$ 
are the gas mass density, energy density, velocity, and pressure
respectively, while $E$, ${\bf F}$, and
${\mathsf P}$ are the radiation energy
density, momentum density or flux, and pressure tensor, respectively.
Only absorptive opacity is included in the Planck mean opacity
$\kappa_P$ and the energy mean opacity $\kappa_E$, but scattering
is added to absorption in the flux-mean total opacity $\kappa$.
Energy injection is permitted via the function $\eta$ (see later discussion).
The orbital frequency, $\Omega = \sqrt{GM/r_o^3}$, is evaluated at the 
central radius $r_o$.
The disk is assumed to be azimuthally symmetric, and shearing-box coordinates 
(the Hill potential) are used to describe the gravity (the $\Omega$-dependent
terms in equation~2 all arise from this approximation to the potential
in rotating coordinates).

    The mass-continuity and Navier-Stokes equations are advanced in time
using operator-splitting.  The radiation transport problem is solved
implicitly using the approximation of flux-limited diffusion (Turner \& Stone 2001).  
The code assumes a gas equation of state with $e$, the gas energy density,
evolved adiabatically, but including absorption and emission of radiation.
The gas pressure is defined as $p_{gas} = (\gamma-1)e$, with $\gamma = 5/3$. 

    In all our simulations, the problem area was a radial segment of a
geometrically-thin, optically-thick, radiation-dominated accretion disk.
In terms of the disk height, $H = F_o \kappa_F/(c\Omega^2)$ as predicted 
by the SS
equilibrium, where $F_o=3\Omega^2\dot M/(8\pi)$ is the radiation flux at 
the top of the disk, $c$ is the speed of light, we simulated the 
dynamics in a region stretching in the
vertical direction from $z=-2 H$ to $z = 2 H$, and in the 
(cylindrical) radial direction from $r = r_o - 2 H$ to $r = r_o + 2H$.
Two values of central radius $r_o$ were chosen, $100r_g$ and $200r_g$,
where $r_g = GM/c^2$.
The black hole mass was $10^8 M_\odot$, and we set the accretion rate
at $\dot M = (3-10)\times 10^{25}$~gm~s$^{-1}$, appropriate for bright active
galactic nuclei shining at (20-60)\% of the Eddington limit.

The initial condition in every case 
was a slightly-modified SS equilibrium.  Because that equilibrium 
becomes ill-defined at the disk surface (where it predicts that the gas density 
falls discontinuously to zero), we constructed a more complete version of the
same equilibrium allowing for the small amount of gas pressure support
that exists at the top.  In this initial equilibrium (but not in our
simulations), we assume that $T_{gas}=T_{rad}$ locally throughout
the disk.  Fixing the flux, $F_o$, column density, $\Sigma$, and orbital
frequency, $\Omega$, we used a shooting method to solve simultaneously
the hydrostatic equilibrium equation,
\begin{equation}
-{d(p_{gas}+p_{rad})\over dz}= \rho \Omega^2 z,
\end{equation}
the radiative equilibrium equation,
\begin{equation}
{dF\over dz} = {2 F_o \rho \over \Sigma},
\end{equation}
and the radiation diffusion equation,
\begin{equation}
F=-{c \over (\kappa \rho + \alpha_o) }{d p_{rad} \over dz}.
\end{equation}
We introduced an scattering opacity floor, $\alpha_o$, to
prevent the opacity from ever becoming small enough to undercut the
validity of the diffusion approximation (see further discussion below).
The gas density in this solution is very nearly
constant from $z=0$ almost to $z=H$; starting from just below $z=H$,
it falls steeply, but not discontinuously, as $z$ increases.  
The radial and vertical velocity components were set to zero initially, while 
the azimuthal velocity components were set to the shearing sheet value,
$-1.5\Omega(r-r_o)$.

     The column density of the accretion disk at the start of the
simulation is chosen to be the value in the the SS equilibrium.  Ignoring
relativistic correction factors and the correction factor accounting for
the outward angular momentum flux, it is
\begin{equation}\label{sigmasoln}
\Sigma = {4 \over 3\kappa\alpha} {\dot m}^{-1} x^{3/2},
\end{equation}
where $\dot m$ is the accretion rate in Eddington units
(for unit efficiency), $\alpha$ is the ratio of the viscous stress to
the pressure, $\Sigma$ has units gm~cm$^{-2}$, and
$x = rc^2/GM$.
In all but one special case, we fixed $\alpha$ at 0.01 for computing
the surface mass density at the start of the simulation. 

At the top and bottom edges ($z = \pm 2H$), the boundary condition was 
chosen to be outflow; that is, fluid quantities in the cells adjacent to the
problem area were set equal to the boundary values. 
In addition, anywhere we set an outflow boundary condition we
required the velocity to be outwards; if not, it was set to zero.
In the radial direction, 
periodic boundary conditions were assumed in all quantities except $v_\phi$,
for which periodicity was enforced for the quantity $v_\phi - v_{Keplerian}$.

The scattering opacity was taken to be the electron scattering value
$\kappa_{es}=0.4 $cm$^2$~gm$^{-1}$, with certain exceptions.  When regions of
the simulation zone were optically thin, we found that the diffusion 
routine took many steps to converge, greatly slowing progress of the
code.  In addition, the outflow boundary condition for optically-thin
flux-limited diffusion was impossible to implement due to a steep
density gradient and uniform radiation field near the boundary where 
the disk has very low density but is nonetheless gas pressure supported.
To avoid these difficulties associated with optically-thin radiation 
transfer, an opacity floor was chosen such that the scattering optical depth across 
each cell was at least unity ($\alpha_o$ was $2\times 10^{-12}$cm$^{-1}$
for simulation 1 and $2\times 10^{-13}$ cm$^{-1}$ for simulation
1).  This adjustment of the opacity affected
about half of the simulation region, but a much smaller fraction of the
total mass.  It caused the radiative flux through 
the upper and lower boundaries to be carried advectively with the fluid
rather than by diffusion relative to the fluid; this forced outflow through
the outer boundaries, maintaining consistency with our outflow
boundary condition.  Over a characteristic thermal time for the
Shakura-Sunyaev equilibrium
\begin{equation}
t_{thermal} \equiv F_o^{-1}\int_0^H (E+e) dz  = 
H\kappa\Sigma/c \simeq 4/(\Omega \alpha), 
\end{equation}
only a small fraction ($\la 1$\%) of the mass within the simulation region 
was lost due to outflow.
The absorption opacity was assumed to be purely bremsstrahlung: 
$\kappa_{abs} \rho= 10^{52} \rho^{11/2}e^{-7/2} {\rm cm}^{-1}$, where $\rho$
is the mass density and $e$ is the gas energy density.

   Without creation of new photons, flux lost out of the top of the disk
would ultimately deplete the disk of radiation.  In a real disk, new photons
are created in a way that depends on the local gas density, temperature,
and magnetic field.  Here we wish merely to model the radiation process
in a phenomenological fashion.  To this end, we simply call the local
radiation rate $\eta$ (units of erg~cm$^{-3}$~s$^{-1}$) and define it
by either of two prescriptions: in one case, $\eta \propto \rho$, while in
the other $\eta$ is
proportional to the total pressure averaged over the simulation region, but
locally proportional to the gas density within any given cell.  The former 
choice was meant to mimic the assumption underlying the SS
vertical equilibrium.  The latter choice is based, of course, on the
thought that the stress is proportional to the total pressure.  

    Ultimately, the energy for these new photons comes from dissipation of
gas motions (and, in a real disk, resistive dissipation of magnetic field
energy).  However, there is no clear-cut way to connect photon creation
to local viscosity.  For example, if, as seems likely, most of the angular
momentum transport in disks is due not to something that behaves like
viscosity, but rather to MHD turbulence (Balbus \& Hawley 1998), there is no
simple, local connection between stress and heating, much less between
stress and radiation.  Consequently, we do not place any
viscous counterpart to the radiation creation term $\eta$ in the gas momentum
and energy equations.

    The final item to be noted in specifying the simulations is the resolution.
The results we present were done with a $128\times 128$ grid, but we reran part 
of simulation 1 with resolutions of $64\times 64$ and $256\times 256$ to check
for convergence.   In those cases in which the disk collapsed as a result
of thermal instability, we stopped the simulation when nearly all the
disk mass was contained within then central 1/4 of the vertical coordinate.  
At that point, the central half (in both radial and vertical directions)
was rebinned to twice the resolution, and the simulation was then restarted.

  The simulations and their characteristics are listed in table \ref{tab1}.
The three parameters varied were the radius, $\dot M$, and resolution;\
the radius and $\dot M$ change the optical depth as described by
equation \ref{sigmasoln},
and also change the ratio of $p_{rad}/p_{gas}$.  
Each simulation consisted of two successive phases:  

A) The disk was first allowed to come to a statistically
steady equilibrium modified by convection with the integrated heating rate 
fixed globally, but scaled locally with the density, $\eta= F_o \rho/\Sigma$. 
This readjustment took several thermal times. 

B)  The heating rate was then allowed to vary in proportion to the average 
total pressure (``$\alpha$'' prescription) to search for possible thermal 
instability.  In this case 
\begin{equation}\label{alphaeq}
\eta=  {F_o\rho\langle p\rangle \over \Sigma \langle p_o \rangle},
\end{equation}
where $\langle p_o\rangle$ is the volume averaged pressure at the 
end of Phase A.

\bigskip
\centerline{Table 1: Simulation Parameters}
\small{
\begin{center}\label{tab1}
\begin{tabular}{|c|c|c|c|c|c|c|c|}
\hline
N &$\dot M$(g/s)& $r/r_g$ &Res.&4H (cm) &$\phi_{rad}$
&$\phi_{gas}$ &$\alpha_o$ \nl
\hline
1 & 3.E25 &  100 & 128$^2$ & 2.4E13 &  95 & 0.50 & 2.E-12\nl
2 & 1.E26 &  200 & 128$^2$ & 1.6E14 & 160 & 0.95 & 2.E-13\nl
\hline
\end{tabular}
\end{center}
}
\section{Readjustment by Convection}

In this section we discuss the results from Phase A of the simulations.

\subsection{Approximate Analytic Description of the Time-Steady State}

   To provide a context for the simulation results, we begin by discussing
how one might find approximate analytic solutions for the vertical structure
of radiation-dominated disks.  Because the inflow time is very long compared
to the dynamical time, $t_{dyn} = 2\pi\Omega^{-1}$, one might expect the disks 
to be in hydrostatic equilibrium.   In the absence of convection, one would also 
expect the heat flux to be carried by photon diffusion.

      On the other hand, when convection is active, one might estimate the
heat flux in terms of mixing-length theory (e.g., as described in Clayton 1968).
If one could define an effective mixing-length $l$, the fraction of the heat
carried by convection is
\begin{equation}
\label{mixflux}
F_{conv}/F_o = 13 (z/h)^{-1/2} (l/h)^2 (\alpha/0.01) (h\Delta\nabla\ln T)^{3/2},
\end{equation}
where $\Delta\nabla\ln T$ is the difference between the true logarithmic
temperature gradient and the adiabatic temperature gradient.  To the
degree that convection is efficient, entropy gradients are erased, so that
the disk becomes nearly isentropic.

    Thus, one way to approximately determine the structure of these disks is to
require all three of these conditions: hydrostatic balance, isentropy, and
photon diffusive equilibrium (or at least that photon diffusion carry a
specified fraction of the heat).  However, this is impossible because any two
of these three conditions
suffice to determine the structure of the disk.  For example, SS applied 
the two conditions of hydrostatic equilibrium and photon diffusive equilibrium, 
without reference to the condition of isentropy (in fact, their equilibrium, 
as we have already remarked, has a strongly unstable entropy gradient).  In 
this solution, the density is constant with altitude up to $H$, and drops 
sharply to zero for $z > H$.  Thus, the mass-weighted average height of the
disk is $\langle z\rangle \equiv \int_0^H \rho z dz / 
\int_0^H \rho dz \sim H/2$.

Similarly, BKB assumed hydrostatic balance and isentropy, but made
no assumption with regard to radiative balance.  If the turbulence is
subsonic, then this approximation is appropriate since the turbulent pressure
will be much smaller than the radiation pressure.  Then, the diffusive
flux is simply 
\begin{equation}
F_{dif}=F_o z/H.
\end{equation}
The density distribution they found is:
\begin{equation}
\rho = \rho_c \left[1-\left({z\over H}\right)^2\right]^3,
\end{equation}
where $\rho_c$ is the central density, given by
\begin{equation}\label{rhoc}
\rho_c = {35 \Sigma \over 32 H}.
\end{equation}
The mass-weighted average height of this disk is $\langle z \rangle = 35H/128 $,
 about half that of the
SS solution.  The specific entropy, $s = {4\over3}a^{1/4}E^{3/4}/\rho$, is found 
to be
\begin{equation}\label{enteq}
s = \left({2\over 105^{1/4}}\right)\left[{a \kappa^7 F_o^7 \over \Sigma \Omega^8 
c^7}\right]^{1/4}.
\end{equation}
In this case, if the dissipation is proportional to density, then the 
convective flux is 
\begin{equation}
F_{con} = {35\over16}F_o \zeta\left({19\over35}-\zeta^2+{3\over5}\zeta^4-\zeta^6/7\right)
\end{equation}
where $\zeta=z/H$.  Thus, for small $\zeta$, the ratio of the convective
flux to the diffusive flux is simply $19/16$;  i.e. the convective flux
is always comparable to the diffusive flux.  At $\zeta=1$, $F_{con}=0$
so all the radiation escapes diffusively.
The volume-integrated radiation energy in the disk is reduced by a factor of 3
with respect to the SS solution, so the thermal timescale is decreased
by the same factor.

Alternatively, one
could also insist that the specific entropy is constant and that the disk
be in radiative balance but pay no attention to hydrostatic equilibrium
(this assumes that the convective flux is small compared to the diffusive flux). 
In that case, if the dissipation is proportional to density, then the density 
profile would be
\begin{equation}\label{isenrad}
\rho = \rho_c {\rm cn}^3\left[{\sqrt{2}\rho_c z\over\Sigma}
\Biggr|{1\over 2}\right],
\end{equation}
where ${\rm cn}$ is a Jacobi elliptic function and $\rho_c$ is the central 
density, given by
\begin{equation}\label{rhoc2}
\rho_c = {\sqrt{\pi}\Gamma(1/4)\Sigma\over 4 \Gamma(3/4) H}.
\end{equation}
The entropy turns out to be identical to equation \ref{enteq}, except for
a different numerical factor in front: 
$4(4/3)^{1/4}\Gamma(3/4)/\Gamma(1/4)/\sqrt{\pi}$,
which is 1.3 times greater than in the BKB solution.
The average radiation energy density in this solution is 4 times smaller
than in the SS solution.
A disk obeying these assumptions would be out of hydrostatic balance by
$\rho_c \Omega^2 H (\sqrt{2} {\rm dn}(u|1/2) {\rm sn}(u|1/2) - z/H)$, where
$u=\sqrt{2}\rho_c z/\Sigma$, and dn, sn are Jacobi elliptic functions.
For hydrostatic equilibrium to exist, turbulence must provide this deficit.
The quantity in parentheses can be approximated by $\sim 0.4\sin{\pi 
(z/H)^{4/5}}$,
so the turbulent velocity must be of order the sound speed, $\sim H \Omega$.

    Thus, at most two of these three plausible conditions can be satisfied
exactly.  On the other hand, it is possible to approximately satisfy all
three if small departures are allowed for each one.  Shakura et al. (1978)
tried to modify both the hydrostatic balance and energy transport equations
to allow for small departures (specifically, the kinetic energy contribution
in the hydrostatic equilibrium equation and convective heat transport in the
energy equation).  However, they were able to find an analytic solution only
for the special case in which the rms Mach number ${\cal M} = 2/3$.  When
that is the case, the density distribution is
\begin{equation}
\rho=\rho_c \left[1-\left({z\over H}\right)^2\right],
\end{equation}
where $\rho_c = 3\Sigma/(4H)$.
Unfortunately, there is no particular reason to
expect that ${\cal M}$ takes this value (in fact, the simulations we
performed indicated that it is smaller by an order of magnitude).
Consequently, it is necessary to solve the equations of motion numerically
in order to find the true compromise between these three conditions that
is found in real disks.

\subsection{Dynamical readjustment due to convection}

   We are now ready to report the results of our simulations.

   Within the body of the disk, the timescale for thermal equilibration of
the radiation and the gas is the shortest timescale; it is a fraction $<1/20$
of a dynamical timescale.  Consequently, deep inside the disk the gas
and radiation temperatures quickly become equal.
Near the surface, the timescale for radiation to
diffuse outwards is comparable to or shorter than the thermal equilibration
timescale, so in those regions the gas temperature is anywhere from 10 to $10^3$
times greater than the radiation temperature.

Though our initial condition satisfies hydrostatic equilibrium, it is 
convectively unstable.  Pietrini \& Krolik (2000) showed
that in the WKB approximation (i.e., $k H \gg 1$) the most rapidly
growing modes in the linear regime are those whose wave-vectors are nearly
horizontal, and that the growth rate for these modes is
\begin{equation} \label{eqconv}
\Omega_{conv}  = \sqrt{3\over 2}\Omega\left({z\over H}\right)
\left[1 - \left({z\over H}\right)^2\right]^{-1/2},
\end{equation}
almost independent of wavelength 
when $\kappa\Sigma \gg 1$ (note the typographical error of a factor of
2 in equation 20 of Pietrini \& Krolik).  That is, they predicted that
convective motions in 
this equilibrium will grow most rapidly near the surface of the disk, and that
the basic growth timescale is the dynamical time.  Our simulations confirmed
this prediction quantitatively for $z/H \la 1/2$;  for example, at 
0.4H the analytic growth time is 9$\times 10^5$s, while the measured
growth time in simulation 1 is $10^6$s.   However, near the disk
surface, the WKB approximation breaks down because there are sharp gradients
in density and pressure.  As a result, the numerical growth rate we
find is slower than the analytic value by a factor of a few near $z\sim H$.

   In the non-linear regime, growth is most rapid for longer wavelengths,
much as in the case of the Rayleigh-Taylor instability (Garabedian 1957).
Horizontal density modulations near the top of the simulated disks with
wavelengths of order $H/10$ grow within a few dynamical times of the start
of the simulation (Figure \ref{fig1}).  Overdense blobs fall, become 
Kelvin-Helmholtz
unstable, and form an inverse-mushroom shape.  The disk rapidly becomes
turbulent, mixing high and low entropy regions, and then smooths out a bit,
while steady convective motions continue.

\vskip 2mm
\hbox{~}
\centerline{\psfig{file=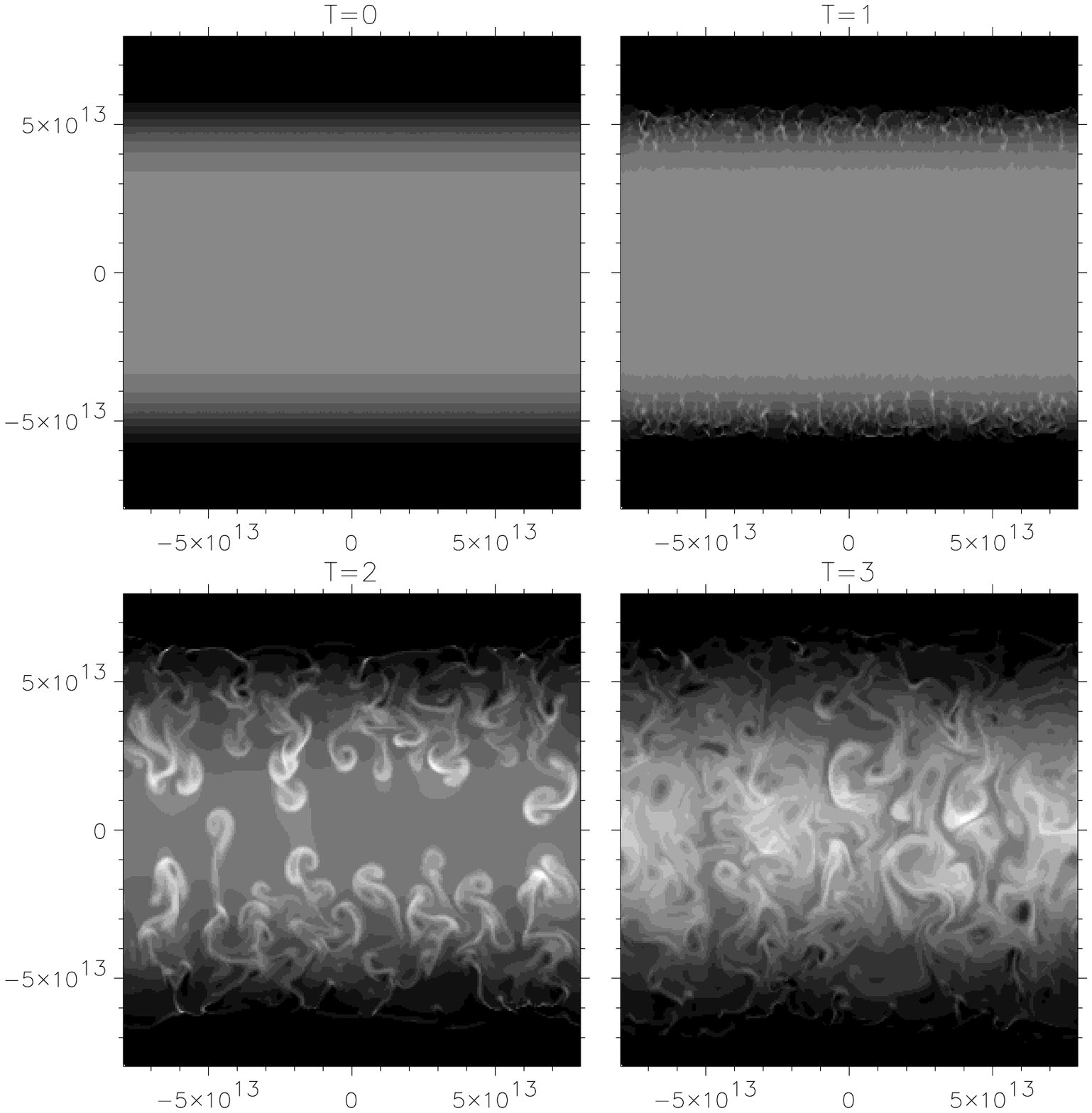,width=3.5in}} 
\noindent{
\scriptsize \addtolength{\baselineskip}{-3pt}
\vskip 1mm
\figcaption{
The density distribution in simulation 1 (at 256$^2$ resolution) at 
times $t/t_{dyn} = 0$, 1, 2, and 3.  The grey scale is linear from 
$\rho = 0$ (black) to $6\times 10^{-9}$~gm~cm$^{-3}$ (white).  The 
horizontal axis is radial distance from the central radius of the simulation,
the vertical axis is altitude from the disk midplane; both are in cm.
\label{fig1}}
\vskip 3mm
\addtolength{\baselineskip}{3pt}
}

    After $\sim$ 90 dynamical times, the disk achieves an approximate 
statistical steady-state.  Convection continues to occur, creating 
small-amplitude fluctuations, but average properties remain very nearly
constant.  The outgoing flux at the top of the box fluctuates about
the mean heating rate, evidence that the simulation has reached a steady state.
Several stable convective cells are formed, in which low entropy 
sinking regions of size $\sim H$ are surrounded by upwelling plumes with twice 
the entropy (Figure \ref{figent}; to be precise, at any given height within
the body of the disk, the specific entropy in the plumes is roughly twice
the lowest specific entropy elsewhere in the disk at that height).
The plumes are not completely resolved since at their narrowest in
the midplane, they are only spanned by 2 pixels.  If the heating rate is proportional
to the local density, mean conditions in the disk remain constant for the
duration of Phase A of the simulation, 11 initial thermal times, or 180 dynamical times.
In contrast to the ``square-wave" density profile of the initial equilibrium,
the mean density profile exhibits a smooth fall from the midplane out to
the top.  The mass-weighted average height of the disk changes by a factor of
$\sim 2$ 
as convection carries more flux outwards, reducing the
radiation pressure, causing the disk to collapse.  The reduced height
further reduces the thermal time, causing more flux to escape, and further
collapse until radiation pressure supports the disk again. At the end of the 
readjustment process, the volume-integrated radiation energy is reduced relative to 
what it was in the initial equilibrium by a factor of 4.4 and 4.6 
in simulations 1 and 2, respectively.  This factor is slightly larger
than the factor of 3 predicted by the analytic BKB scaling, but is close to
the factor of 4 predicted by equation \ref{isenrad}.
We reran simulation 1 at lower ($64^2$) and higher resolution (256$^2$) to check 
for convergence.  We found that the radiation energy density in the $64^2$
simulation disagreed by 20\% at the end of phase A, demonstrating that
at this resolution the simulation was not converged.  At higher resolution, 
$256^2$, we were only able to run the simulation for about a thermal time since 
the run time scales as $N^4$, but we found that the difference 
between the total radiation energy density of the 128$^2$ and 256$^2$ simulations 
had a standard deviation of only 0.6\%, indicating that the 128$^2$ simulation 
was indeed converged.  We utilize this higher resolution run when studying
the initial convective collapse (e.g. Figure \ref{fig1}).

\vskip 2mm
\hbox{~}
\centerline{\psfig{file=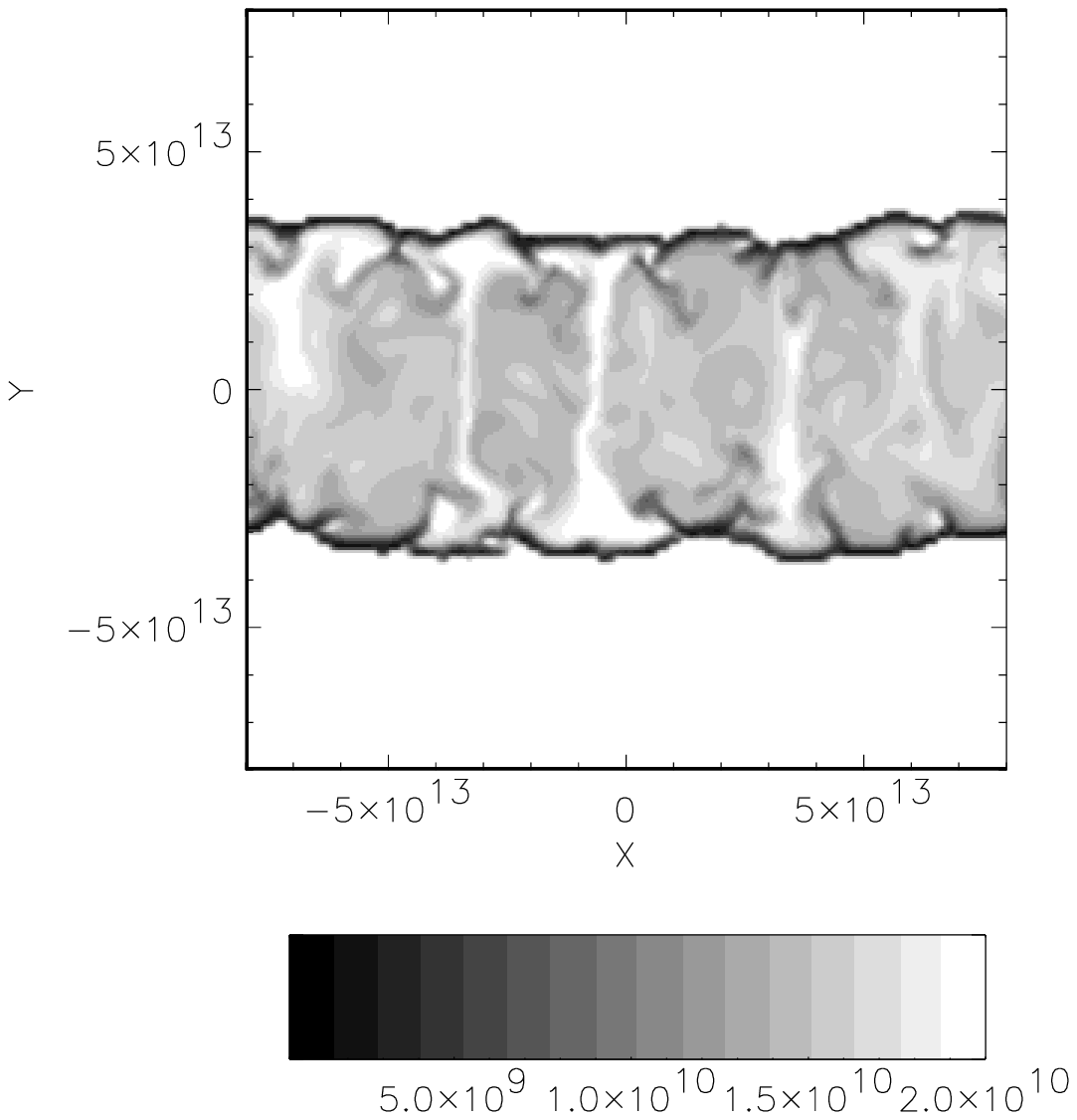,width=3.5in}} 
\noindent{
\scriptsize \addtolength{\baselineskip}{-3pt}
\vskip 1mm
\figcaption{
The specific radiation entropy distribution in simulation 1 (resolution 128$^2$)
in units of cm$^2$~s$^{-2}$~K$^{-1}$ at $t=97 t_{dyn}$.  Axes are as in Fig. 1;
regions above the maximum of the color scale are white.
\label{figent}}
\vskip 3mm
\addtolength{\baselineskip}{3pt}
}
   The mean time-steady density profile of simulation 1, Phase A, is shown 
in Figure \ref{fig3}, where it is contrasted with the various approximate 
analytic solutions discussed in the previous subsection.  
This distribution lies somewhere 
between the BKB and equation \ref{isenrad} solutions.  These analytic solutions 
break down near the surface where, in actuality, gas pressure support
becomes important and LTE no longer holds.  We do not find any dependence
of the inner disk structure on the opacity floor (we increased it by a factor
of 2 and found the disk structure was identical).
The Shakura et al. (1978) solution, with column density equal
to the simulation result, deviates significantly from the simulation mean.
\vskip 2mm
\hbox{~}
\centerline{\psfig{file=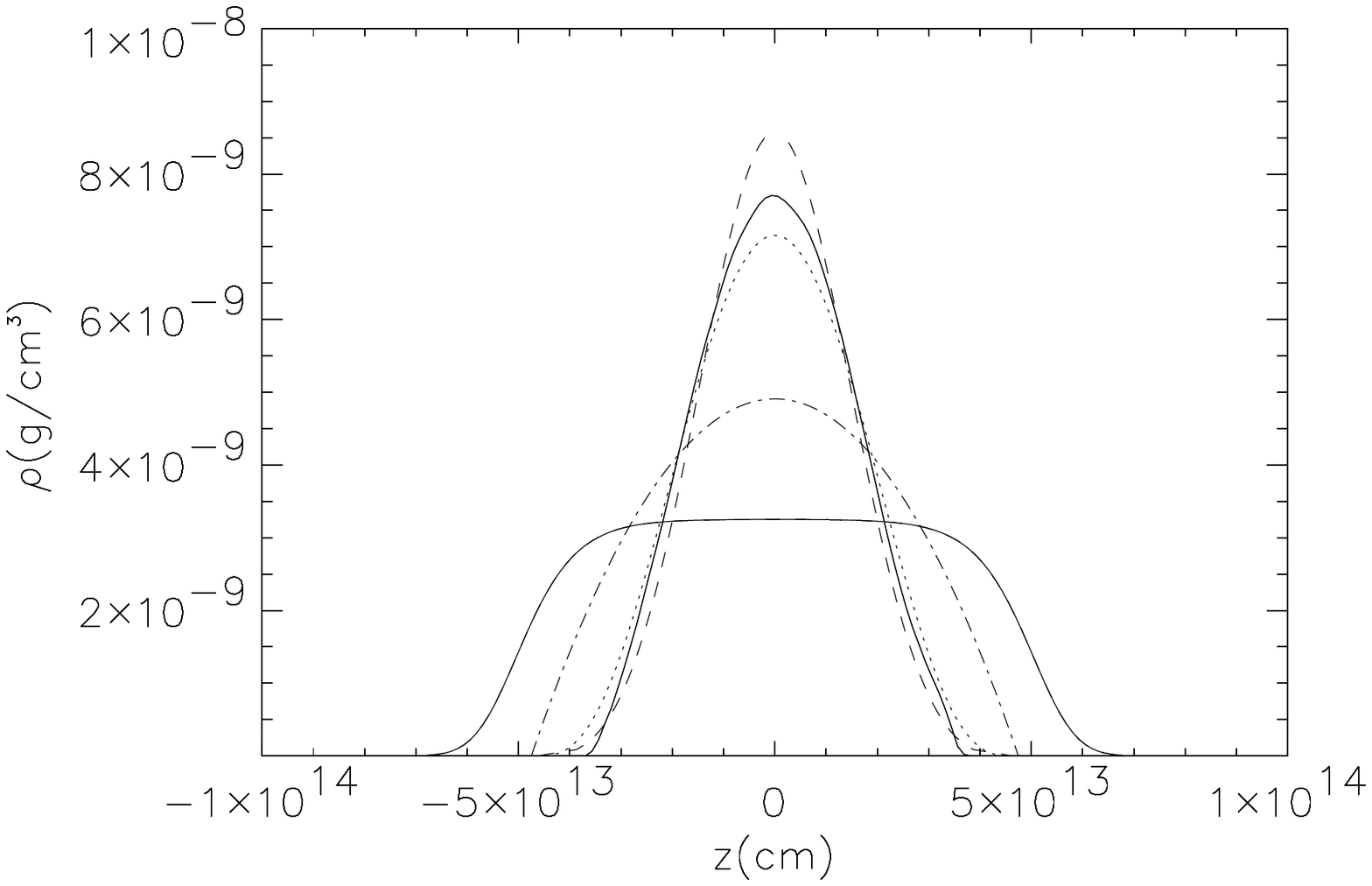,width=3.5in}} 
\noindent{
\scriptsize \addtolength{\baselineskip}{-3pt}
\vskip 1mm
\figcaption{Density distribution at $t = 0$ (flat-topped solid line) and 
$t=90t_{dyn}$ 
in simulation 1 (bell-shaped solid line) compared to the prediction of 
BKB (dotted line), 
equation \ref{isenrad} (dashed line), and Shakura et al. (1978)
(dashed-dot line).
\label{fig3}}
\vskip 3mm
\addtolength{\baselineskip}{3pt}
}

\subsection{Heat Transport} \label{heat}

     The most dramatic feature in Figure~\ref{figent} is the sharp contrast
in specific entropy between rising and sinking regions.  Nonetheless, as
one might expect from the nonlinear development of an instability that
feeds on vertical entropy gradients, convection does an excellent job of
smoothing out the radially-averaged specific entropy profile
(Figure \ref{fig2a}).  Within a few dynamical times, the radially-averaged
specific entropy in the disk becomes very nearly constant
for $z \la 0.8 H$; this (on-average) isentropic region contains
98\% of the disk mass.  Convection is so efficient that this is accomplished
with relatively small amplitude motions: the rms Mach number in
simulation 1 is ${\cal M} \sim 0.1$, vindicating the quasi-hydrostatic
approximation made in the BKB solution. 

    Although the radially-averaged density distribution in the disk
is close to that predicted by the two isentropic models (see Fig.~\ref{fig3}),
the actual value of the mean specific entropy in the
disk is about 25\% smaller than the analytic prediction (equation \ref{enteq}). 
This departure may be due to the fact that the entropy is set by the
requirement of radiative diffusion matching the total flux at the top
of the disk.
In the simulations, the density gradient near the disk surface is somewhat
different than that predicted by the analytic models
because turbulent support becomes comparable to radiation support 
(${\cal M} \sim 1$).  As a result, the entropy necessary to maintain
the flux can be slightly different.

     This strong turbulence near the disk surface finds its ultimate origin
in the reduction in optical depth across convective cells as the density
falls sharply near the disk surface.   When photons can diffuse across
a cell in an eddy turnover time, the radiation does not effectively couple
to the fluid, and the sound speed relevant to fluctuations on those
lengthscales drops to the (much lower) gas sound speed.  Because
photon diffusion prevents the radiation from being compressed along with the
gas, the gas's compressibility increases dramatically.  That permits
large local density enhancements, which lead in turn to rapid cell-sinking and
high speed motions.

     Compressive motions call into play the artificial viscosity employed
in Zeus.  In simulation 1, Phase A, dissipation due to artificial viscosity
comprises 9\% of the total heating rate within the box.
In the immediate vicinity of a shock, the dissipation can
be significant, but averaged over larger volumes and times, it does not
qualitatively alter the disk's thermal properties.

\vskip 2mm
\hbox{~}
\centerline{\psfig{file=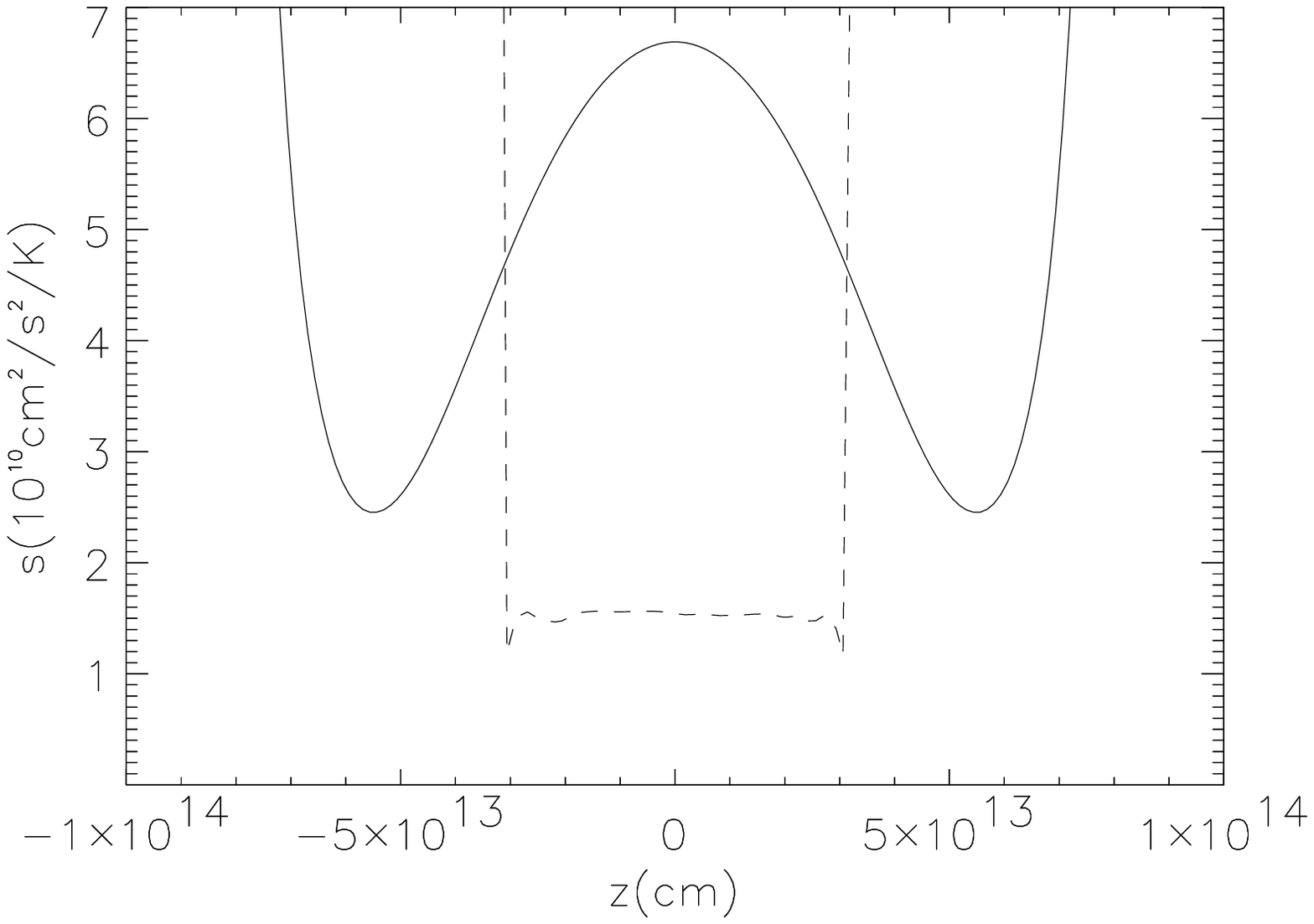,width=3.5in}} 
\noindent{
\scriptsize \addtolength{\baselineskip}{-3pt}
\vskip 1mm
\figcaption{Horizontally-averaged entropy as a function of height at t=0
(solid line), and at $t=95t_{dyn}$ (dashed line) from simulation 1.
\label{fig2a}}
\vskip 3mm
\addtolength{\baselineskip}{3pt}
}

The mean diffusive flux, $F_{dif} = \langle c\partial_z E /
(3\kappa)\rangle $ and 
the convective flux, $F_{con}=\langle v_z (E+e) + \int_0^z dz (\bnabla {\bf 
v}:{\bf P}
+p \bnabla \cdot {\bf v})\rangle $, averaged
radially and in time, are shown in  \ref{fig4}.
The convective flux is comparable to the diffusive flux near the midplane of
the disk, as predicted by the 
BKB solution.   
It is, of course, the additional heat loss due to convection that reduces the
thermal time so much.
In line with intuitive expectation, the diffusive flux rises steadily
with height within the disk until it carries almost all the heat at
the disk surface.  The integrated heating rate and
the total flux are nearly equal at the top of the box, indicating that
the disk is in quasi-equilibrium.

    With these observations in hand, we can evaluate the possible
relevance of estimates made through mixing-length theory.  On the one
hand, the fact that the mean specific entropy is very nearly constant
as a function of altitude is consistent with the usual expectation that
convection very nearly erases any difference between the actual temperature
gradient and the adiabatic temperature gradient.   Moreover, our finding
that the convective and diffusive heat fluxes are similar deep inside the
disk could in principle allow us to turn around equation \ref{mixflux}
and evaluate an effective mixing length.  However, there are both technical
and conceptual problems preventing this.  The technical problem is that the
limited numerical accuracy with which we can measure the mean temperature
gradient will lead to a highly uncertain estimate of the {\it difference}
between it and the very similar adiabatic gradient.  The conceptual
problem is that the large instantaneous
contrasts in specific entropy at fixed altitude demonstrate that the
mixing length picture has questionable validity here because the cell sizes
are comparable to the entire thickness of the disk.

   Above the surface of the disk, the opacity floor that we impose creates
a numerical artifact in the character of heat transport.  With the true
opacity, that region would be optically thin and the heat would be
carried almost exclusively by free-streaming radiation; the opacity floor
causes the heat to be carried advectively.
We believe that this has little consequence for dynamics in the disk
body because the region in which the opacity floor is implemented
contains only $\sim 1\%$ of the disk mass, and because a simulation with
an opacity floor twice as large gave similar results.
\vskip 2mm
\hbox{~}
\centerline{\psfig{file=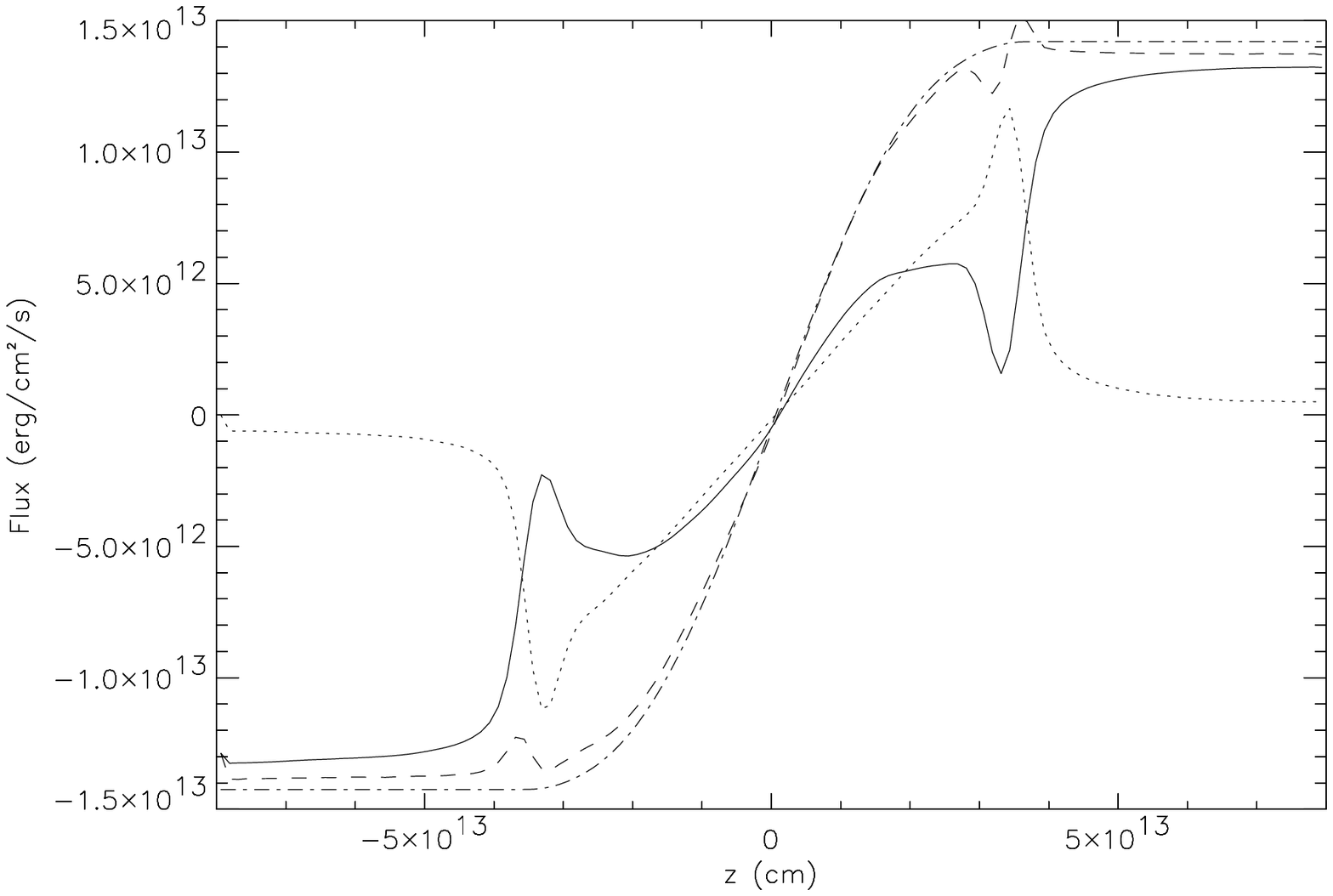,width=3.5in}} 
\noindent{
\scriptsize \addtolength{\baselineskip}{-3pt}
\vskip 1mm
\figcaption{Horizontally-averaged diffusive energy flux (dotted line) compared 
to the convective flux (solid line), total flux (dashed line), and integrated
heating rate (dot-dash line) from $t/t_{dyn}=90$ to 95 in simulation 1.  The 
dominance
of convective heat flux for $|z| \ga 4 \times 10^{13}$ cm is an artifact
of our opacity floor; in real disks the flux at high altitude is 
free-streaming.
\label{fig4}}
\vskip 3mm
\addtolength{\baselineskip}{3pt}
}

\section{Evolution on the Thermal Timescale}

\subsection{Thermal Instability}

One of the important questions motivating this study is whether the thermal
instability predicted on the basis of the $\alpha$-model
actually occurs in radiation-dominated disks.   We began this part of our
study by verifying that our simulation code reproduced this instability in
1-D when the radiation energy generation rate is proportional to the local
radiation energy density.   We increased $\alpha$ to 0.1 in this test (only) to
reduce the thermal timescale, thereby avoiding a long-term drift in the
radiation pressure which occurred only in the 1-D simulations.\footnote{The drift
is an artifact of the opacity floor.  When the radiation flux is purely
diffusive, only the radiation energy gradient matters; consequently, the
general level of radiation energy density is determined only up to an additive
constant.
In real disks, the diffusion approximation breaks down in the optically thin
region, so it is possible to think of this additive constant as determined by the
condition that the flux must be carried by free-streaming radiation at the
top of the box.   However, in our 1-D simulation the flux is diffusive
everywhere due to the opacity floor, so there is nothing in the equations
to prevent numerical drifts in the general level of radiation energy
density.  By contrast, in the 2-D simulations, the flux leaving the box
is carried by advection of radiation, which then fixes the radiation normalization
as in the free-streaming case.}

The growth timescale predicted by Shakura \& Sunyaev (1976)
in the long-wavelength, radiation-dominated limit is
$5(\alpha \Omega)^{-1} = (5/4)t_{therm} \simeq 8 t_{dyn}$, numerically equal to
$2.5\times 10^7$~s for $\alpha=0.1$ and $r/r_g=100$.  Our measured growth time
was 2.5$\times10^7$s, a remarkable confirmation of their analytic prediction.

On the other hand, in 2-D, convection becomes possible, so we now describe
Phase B of the simulations.  To test for thermal instability, the disk was 
first
evolved over several thermal timescales with the heating rate proportional to 
the total mass (Phase A described in the previous section).  We began with this
prescription for the heating rather than the $\alpha$-model for two reasons.
First, we wished to
separate evolution driven by the convective instability from evolution
driven by thermal instability.  Second, as a result of the heat lost by
convection, the total pressure in the disk after convection sets in is smaller
than the initial total pressure.  Consequently, if the heating rate is
set by proportionality to the total pressure, the proportionality constant
that gives thermal balance for the initial disk configuration will not
result in enough heating to maintain thermal balance after the convective
readjustment.   Equation \ref{alphaeq} ensures thermal balance at the beginning
of Phase B.

   At the start of Phase B, we first let the thermal instability grow out of
numerical noise.
Comparing the total radiation energy in the box at the
time the heating prescription was changed to $\alpha$ dissipation to the
time-averaged radiation energy in the simulations with dissipation
proportional to gas density, the integrated numerical fluctuation in
the radiation energy was less than 0.1\%.  In both simulations, as soon as
the heating prescription was changed, departures from equilibrium began
to grow.  The initial growth was well described by an
exponential; in simulation 1, its e-folding time was
$\sim 18 t_{dyn}$, $\simeq 0.9$ thermal times as measured in the
convective equilibrium.   Simulation 2 showed the same growth time,
$\sim 18 t_{dyn}$.  Because the single-zone linear theory of
Shakura \& Sunyaev (1976) was specific to their equilibrium, we cannot
directly compare this growth time with analytic predictions; however,
it is comparable to the thermal time of the collapsed disk, $\sim 16 t_{dyn}$.  
Thus, as predicted by Robertson \& Tayler (1981), convection does not
quench the thermal instability.  However, as we shall see momentarily,
it does significantly change its character.

In both simulations, numerical noise lead to disk collapse.  However,
the pure $\alpha$-model without convection predicts that the instability has the
same growth rate whether the sign of the temperature perturbation is
positive or negative.  To test whether radiation energy growth can also occur,
we reran Phase B with the change that we artificially increased the radiation
energy density by 5\% at the start.  In the $r=100r_g$ case, the disk still 
collapsed on a thermal timescale.  However, in the $r=200r_g$ case, the radiation 
energy density grew on a thermal timescale.  The simulation was stopped after 
most of the mass was lost through outflow from the simulation grid.
We speculate that the positive energy perturbation at $r=100r_g$ was
reversed because the initial growth in radiation energy triggered such
strong convection that soon the disk was losing more heat than it gained. 
Because the primary distinction in physical conditions between the two
cases was that the ratio of radiation to gas pressure was larger by
a factor of 1.7 at $r=200r_g$ than at $r=100r_g$, it is possible that their
contrary fates were related to this fact.  However, we have not been
able to convincingly identify any particular diagnostic that enables us to 
predict which cases will collapse rather than expand despite initially positive
temperature perturbations.

	We also tried the further experiment of
imposing a factor of two increase in the local radiation energy density
at the end of Phase A of the $r=100 r_g$ simulation.  This had
the result of the disk heating and expanding until a significant amount of
mass was lost due to strong outflow.

   On the basis of these numerical experiments, we conclude that
convection biases the development of the thermal instability
in the linear regime so that negative radiation energy density
perturbations are favored.  In some circumstances, 
this bias is strong enough that even when the initial perturbation
is an increase in the radiation energy density, the ultimate
result of the instability is collapse.  However, this is not
always possible, and sufficiently large positive perturbations can overcome
convection and lead to runaway growth in the radiation energy density.
Our simulations, unfortunately, are not capable of following runaway
growth very far because so much of the disk mass is quickly pushed out
of the problem area.  Consequently, we cannot say how these cases
develop in the long-term. 

   In all these numerical experiments we have restricted ourselves to
very simple phenomenological descriptions of the heating and, in the
case of the simulations with imposed perturbations, very simple and
uniform perturbations.  Real disks undoubtedly behave differently.
We regard these simulations therefore as demonstrating that it is possible
for thermal instability to occur after convective readjustment
when the heating rate is roughly described by the $\alpha$-model,
and that in some circumstances, convection can cause a bias toward
collapse.  However, these simulations certainly do not provide the
final answer to the question of whether and how thermal instability
affects physical disks.

\subsection{Final state}

After the initial exponential collapse, the radiation energy density 
approaches a constant value as the disk becomes gas-pressure supported
and thus thermally stable.  This is a consequence of the fact that for
fixed surface mass density $\Sigma$ and fixed stress-parameter $\alpha$,
there exist exactly two thin disk solutions, one radiation-pressure
supported with high accretion rate and flux, and one gas-pressure
supported with low accretion rate and flux (an advection-dominated
solution can also exist, but it is no longer thin).
Figure \ref{fig6} shows the radially averaged
gas density in the final state of simulation 1, phase B, while
figure \ref{fig7} shows the radiation and gas pressures.    Despite
the fact that the gas pressure is is larger within the bulk of the disk,
$p_{rad}$ becomes larger than $p_{gas}$ near the photosphere simply
because the gas density drops precipitously.  The density
weighted mean of $|z|$ is smaller by a factor of 12.5 at the end of 
Phase B than at the beginning of Phase A.  Twice the grid is rebinned
by a factor of two to follow the collapse.  We find that the gas becomes 
strongly clumped in the radial direction filling only $\sim 1/4-1/2$ of the 
simulation region.  Since centrifugal forces balance gravity in the radial 
direction, the clumps remain stable.  In a real disk, we expect 
that viscous stresses will erase this radial stratification.

To illustrate the relationship between the initial radiation-pressure
supported state and the final gas-pressure supported state,
we explicitly find them in terms of the parameter $\phi=p_{rad}/p_{gas}=
({1\over 3}aT^4)/(\rho k_B T/m)$ ($m$ is the average mass per particle). 
Ignoring the radial clumping, we approximate the disk as a single vertical
zone from $z=0$ to $H$, writing the hydrostatic equilibrium equation in the 
form:
\begin{equation}
(1+\phi)p_{gas} = \rho H^2 \Omega^2.
\end{equation}
In the $\alpha$ prescription, the radiation equilibrium equation becomes 
\begin{equation}
{3\over 2} \alpha \Omega (1+\phi) p_{gas} H = {4 c \phi p_{gas}\over \kappa 
\Sigma}.
\end{equation}
Defining $\Sigma = 2\rho H$, we solve for $\phi$:
\begin{equation} \label{phieq}
(1+\phi)^{10} = \phi^6 \gamma_o
\end{equation}
where $\gamma_o$ is defined as
\begin{equation}\label{gameq}
\gamma_o = \left({8c\over 3\alpha\kappa_{es}\Sigma}\right)^7
{2am^4\over 3\Sigma\Omega k_B^4}.
\end{equation}
When radiation pressure dominates, $\phi\gg1$, so  $\phi_{rad} \sim 
\gamma_o^{1/4}$.
When gas pressure dominates, if $\phi \ll 0.1$, then 
$\phi_{gas} \sim \gamma_o^{-1/6}$.
Thus, we expect that $\phi_{gas} \sim \phi_{rad}^{-2/3}$.  If $\phi_{gas} \sim 1$,
then it must be solved for numerically using equation \ref{phieq}.

\vskip 2mm
\hbox{~}
\centerline{\psfig{file=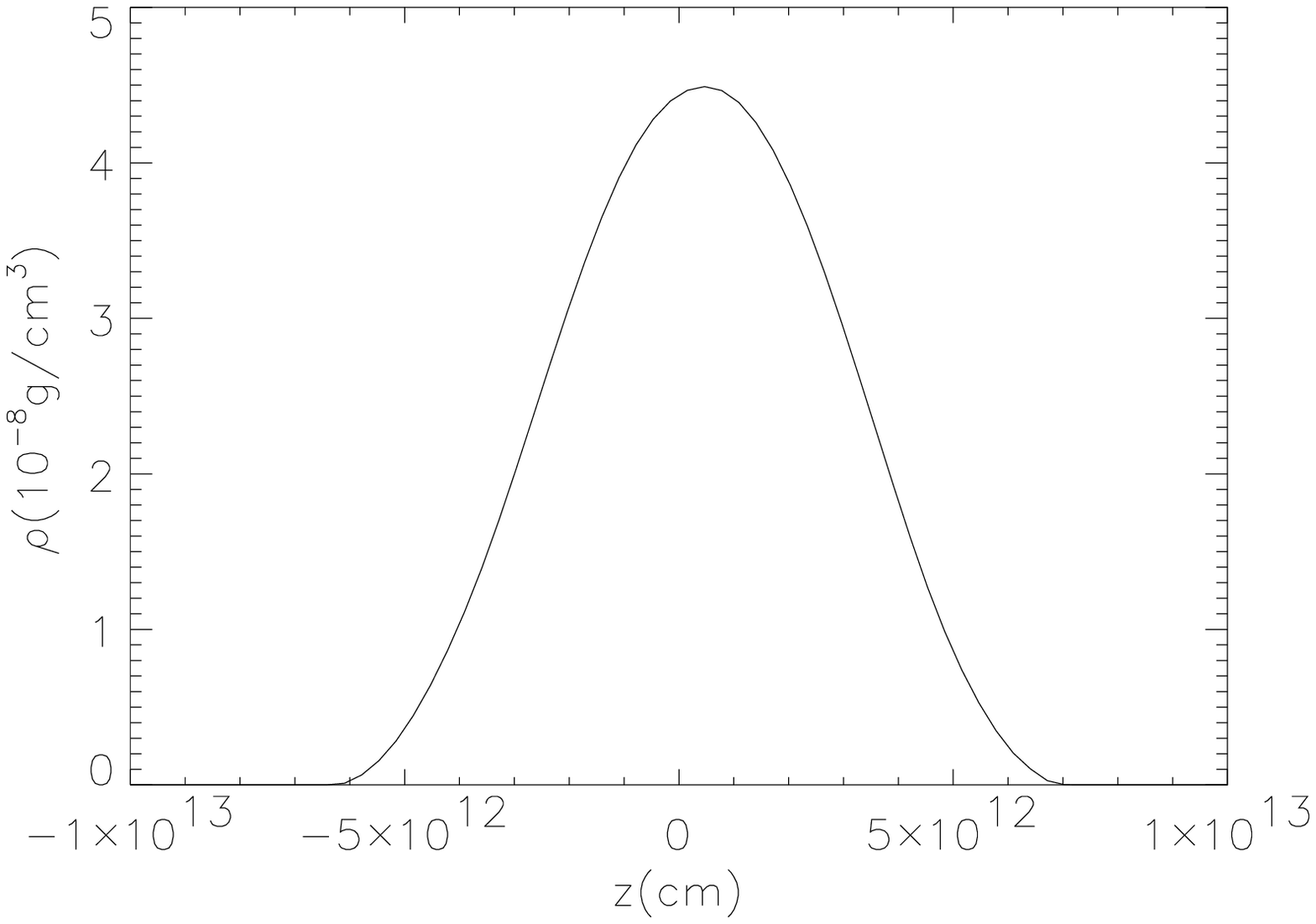,width=3.5in}} 
\noindent{
\scriptsize \addtolength{\baselineskip}{-3pt}
\vskip 1mm
\figcaption{Radially-averaged density profile at the end of simulation
1, phase B.
\label{fig6}}}
\vskip 3mm
\addtolength{\baselineskip}{3pt}

\vskip 2mm
\hbox{~}
\centerline{\psfig{file=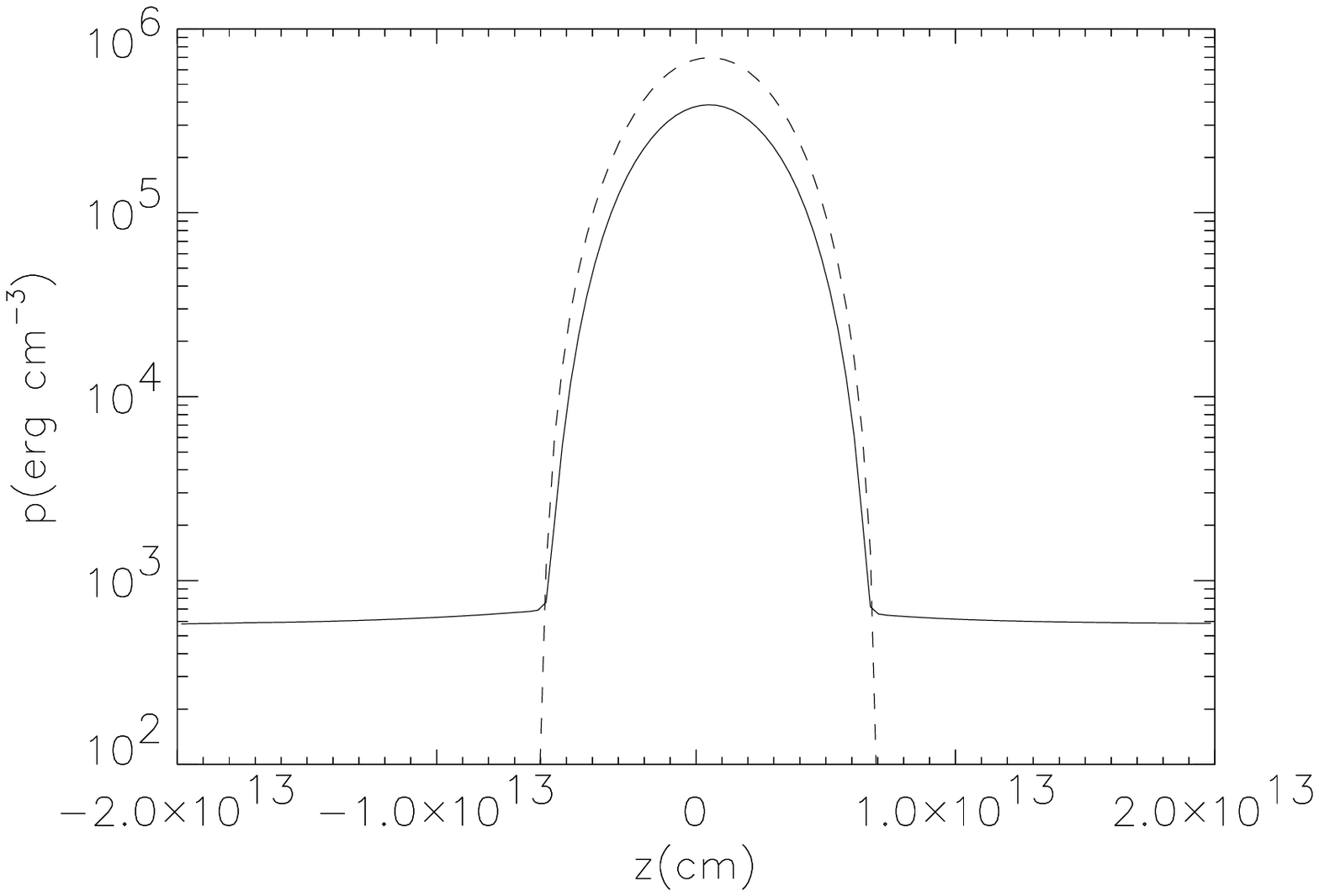,width=3.5in}} 
\noindent{
\scriptsize \addtolength{\baselineskip}{-3pt}
\vskip 1mm
\figcaption{Radially-averaged pressure profiles at the end of simulation
1, phase B.  Solid line is radiation pressure, while dashed line
is gas pressure.
\label{fig7}}}
\vskip 3mm
\addtolength{\baselineskip}{-3pt}

Due to the clumping, this relation is only qualitatively borne out by our 
simulations.  Simulation 1 begins with $\phi_{rad} = 95$ (density-weighted 
average).  As a result of the $\alpha$ renormalization that occurs in order 
to preserve thermal balance when the heating prescription is changed, $\alpha$ 
changes from 0.01 to $\sim$ 0.05 (since the
radiation pressure is reduced by $\sim 1/5$).  With this new value of $\alpha$,
we would expect on the basis of the simple one-zone model just presented that
the final $\phi_{gas}$ after collapse should be $\sim 0.9$; we find 0.54 in 
simulation 1.  Similarly, simulation 2 begins with $\phi = 160$.  The 
effective $\alpha$ again changes from 0.01 to 0.05, leading to a predicted 
final $\phi_{gas} = 0.38$ (from equation \ref{phieq}); we find $\phi_{gas} \simeq 
0.95$.  The column density is artificially increased due to the clumping, so the 
effective $\gamma_o \propto \Sigma^{-8}$ decreases (equation \ref{gameq}), 
leading to a larger $\phi_{gas}$.  The radial clumping is much stronger
in simulation 2 than in simulation 1, which explains the larger discrepancy.
In simulation 2, $p_{rad} \sim p_{gas}$ in the final state, so weak convection 
is still present.  

\section{Conclusions}

\subsection{Evolution on the dynamical time-scale}

We have shown that in a very short time (of order the dynamical time),
convection modifies the structure of geometrically thin radiation-dominated
disks.  Although there are order unity contrasts in specific entropy
at fixed altitude $z$, the radially-averaged entropy becomes
very nearly constant with height as a result of convective mixing.
The mean density profile that emerges is close to the one predicted in
the analytic solution of BKB.  As likewise predicted by that analytic
solution, the convective flux near the disk midplane is
comparable to the diffusive flux.  More rapid upward heat flux (for
the same dissipation rate) results in smaller mean radiation pressure,
weaker support against gravity, and therefore greater mean density.
The decreased pressure may cause a further decrease in the viscous 
stress, resulting in a higher surface mass density and thus lower 
radiation to gas pressure ratio.   One consequence may be that
the region of thermal instability is reduced in size by a factor
of a few, changing the accretion rate at which near-Eddington disks 
are subject to thermal instability.

\subsection{Evolution on the thermal timescale}

    We have also shown that evolution on the thermal timescale is
quite sensitive to the specific character of the dissipation prescription.
Dissipation proportional to density leaves the disk in a long-term
statistical equilibrium in which convection and radiative diffusion
carry almost equal amounts of energy.

    On the other hand, dissipation proportional to vertically-integrated
pressure is subject to the thermal instability first pointed out by Shakura \&
Sunyaev (1976), but subject to a significant modification due to
convection: it is biased toward collapse, rather than runaway
expansion.

   The result of thermal collapse is a new gas pressure-dominated
equilibrium at the old column density.  The pressure in this new equilibrium
is much smaller than in the original one.  Because the stress such a disk
is capable of exerting (assuming the $\alpha$ model still holds at least
approximately, as extensive MHD simulations indicate: Balbus \& Hawley
1998) is much smaller than in the initial state, the mass accretion rate
through such a disk is much smaller than the initial value.  Consequently,
if outer regions of the disk continue to pass matter inward at the
same rate, mass must build up in the collapsed portions of the disk.

   The long-term evolution (i.e., on the inflow timescale) of this situation
is difficult to predict.  We speculate that the column density will continue 
to build up until thermal instability sets in at $p_{rad} \sim p_{gas}$. 
The specific point at which this occurs may be modified by convective heat 
transport.  At this point, the disk no longer has a gas-pressure supported 
equilibrium available for fixed column density, so the disk may heat up, 
joining the advection-dominated ``slim disk" solution, leading to limit
cycle behavior on the inflow timescale (Abramowicz et al. 1988), or 
convection may lead to episodic release of energy on the dynamical timescale, 
matching on average the energy dissipated within.

\subsection{Future improvements}

We have made several assumptions to simplify our calculations
that can be tested with future work.  First, we have assumed
that flux-limited diffusion is an appropriate description of the
radiation field.  Since our simulations were optically thick everywhere,
this may be appropriate, but needs to be checked with a more accurate
computation, for example using the variable Eddington factor method
described in Stone, Mihalas, \& Norman (1992).

Second, we have assumed simple dependences of the heating rate on local
disk conditions.  Actual disks may have a heating rate which depends on
global disk parameters, e.g. height or radius.  If a large fraction of the 
magnetic energy is carried to the corona, then the disk may never become
radiation-pressure supported (e.g., as suggested by Svensson \& Zdziarski
1994).  Simulations of disk annuli indicate that a significant fraction
of the magnetic energy generated within an accretion disk can be
carried to the corona (Miller \& Stone 2000), but not necessarily
enough to eliminate a radiation-pressure dominated disk.  These simulations
need to be extended to include radiation and a physical equation of 
state to determine the magnitude of the transported energy.
Because disks with dissipation primarily at the surface are
gas pressure-supported, they are subject to neither convective nor
thermal instability.

We believe that the problems of the previous paragraph may best be addressed 
with 3-D radiation magnetohydrodynamic simulations in which the 
magnetic field strength and dissipation rate are computed self-consistently.
The nature of convective transport as well as the contribution of
turbulence to heat transport (Balbus 2000) may have a large effect 
on our results.   Magnetohydrodynamic turbulence might destroy the
convective plume structure since the magnetorotational instability 
operates on the dynamical timescale.   The damping of turbulence by
radiation diffusion or shocks at the scale $\alpha H$ might
determine the heating profile of the disk (Agol \& Krolik 1998).
If the magnetic stress scales with gas pressure rather than radiation 
pressure, the thermal instability is suppressed (Piran 1978);  whether 
this is the case can be addressed with radiation MHD simulations.
We have purposely limited the scope and detail of our current 
simulations as our neglect of 3-D magnetohydrodynamics will change
the nature of the equilibrium.

\acknowledgements

We thank Paola Pietrini for very useful discussions.  We thank the referee
for useful comments which greatly improved the paper.  The work of NT and JS 
was supported by DOE grant DFG0398DP00215.   The work of EA and JK was 
partially supported by NASA Grant NAG 5-3929 and NSF Grant AST-9616922.
Partial support for EA was provided by NASA through Chandra Postdoctoral
Fellowship grant number PF 0-10013 awarded by the Chandra X-Ray Center,
which is operated by the Smithsonian Astrophysical Observatory for NASA
under contract NAS8-39073.

\end{document}